# The Naked Eye Stars as Data Supporting Galileo's Copernican Views

Christopher M. Graney
Jefferson Community College
1000 Community College Drive
Louisville, Kentucky, USA 40272
www.jefferson.kctcs.edu/faculty/graney
christopher.graney@kctcs.edu

**Abstract**

The aggregate appearance of the naked-eye stars would appear to Galileo to be direct observational support for his ideas about the stars, and indirect observational support for the Copernican theory over the rival Tychonic theory. Brief historical background is also included.

**Introduction**

In the late 16th century Tycho Brahe introduced his geocentric theory, in which the sun and moon circled the Earth while the planets circled the sun.  This was observationally equivalent to the Copernican theory insofar as the sun, moon, and planets were concerned; astronomical observations of these bodies, including Galileo Galilei's famous telescopic observations of them, could not distinguish between the two theories.  Galileo argued that the tides revealed Earth's motion but that was a flawed argument, recognized as such by anti-Copernicans.[1]  In principle, observations of the "fixed" stars could distinguish between the two theories -- for in one Earth was at rest with respect to the stars and in the other it wasn't -- by revealing annual stellar parallax caused by Earth's motion relative to the stars, for example.  Lack of such observations meant Robert Hooke could state in the 1670's that the Tychonic theory had much validity and that arguments regarding Tycho versus Copernicus might go on indefinitely absent detection of parallax.[2]  Observational evidence for Earth's motion arrived in 1728, when James Bradley detected the aberration of starlight due to the relative motion of the Earth and stars, by which time Newtonian physics was providing a framework of understanding that supported the Copernican theory but not the Tychonic one.[3]

    Long before Bradley and Newton, Galileo had strongly backed the Copernican theory.  Coincidentally, his ideas about the fixed stars, followed to their logical implications regarding

the aggregate appearance of the naked-eye stars, lead to the conclusion that said appearance supports the Copernican theory.

**Following Galileo on the Fixed Stars**

Galileo's interpretation of the Copernican theory included something not original to Copernicus -- that the fixed stars were "so many suns"[4] distributed through space so that "some are two or three times as remote as others"[5]. This assumption that the stars are essentially identical to the sun was a prominent and recurring feature in Galileo's thinking.

Galileo believed that with his telescope he could see the physical globes of stars and accurately measure their angular diameters ($\alpha$). He used such measurements, in conjunction with the assumption that the stars were suns, to determine stellar distances:

- He measured (1617) $\alpha$ of the brighter component of the double star Mizar as being 300 times smaller than $\alpha$ of the sun and calculated it to be 300 AU[§] distant[6]

- He argued (1624) that "...the sun's diameter is five hundred times that of an average fixed star; from this it immediately follows that the distance to the stellar region is five hundred times greater than that between us and the sun"[7]

- He stated (1632) that sixth-magnitude stars have $\alpha$ 2160 times smaller than the sun, and therefore lie at a distance of 2160 AU[8]

---

§ *The Earth-Sun distance is 1 AU.*

So for Galileo, the distance $L$ in AU to a star whose angular diameter he measured to be $\alpha$ is $L = \alpha_\odot/\alpha$ ($\alpha_\odot$ is the angular diameter of the sun). The number of stars $N^*$ located within a radius $r$ of Earth must be given by $N^*(r) = 4/3\ \pi\ \rho^*\ r^3$, where $\rho^*$ is the average number density of stars in space; the number of stars with angular diameter of $\alpha$ or greater is then given by

(1) $\qquad N^*(\alpha) = 4/3\ \pi\ \rho^*\ (\alpha_\odot/\alpha)^3$.

Now consider another of Galileo's ideas concerning the stars -- the relationship between a star's $\alpha$ and its naked-eye magnitude $M$. Galileo stated that for stars of $M = 1$, $\alpha = 5''$‡; for stars of $M = 6$, $\alpha = 5/6''$.[9] (See Addendum for more on Galileo's star sizes.) This suggests (Figure 1),

(2) $\qquad \alpha = (5/6)(7 - M)$

so the number of naked eye stars visible from Earth of magnitude $M$ or brighter is (Figure 2)

(3) $\qquad N^*(M) = 288/125\ \pi\ \rho^*\ (\alpha_\odot/(7-M))^3$

**Comparison to Data**

Galileo could use equation 3 as a test of his ideas -- a count of naked eye stars yielding data consistent with equation 3

---

‡ *5″ = 5 seconds of arc = 5/3600 degree*

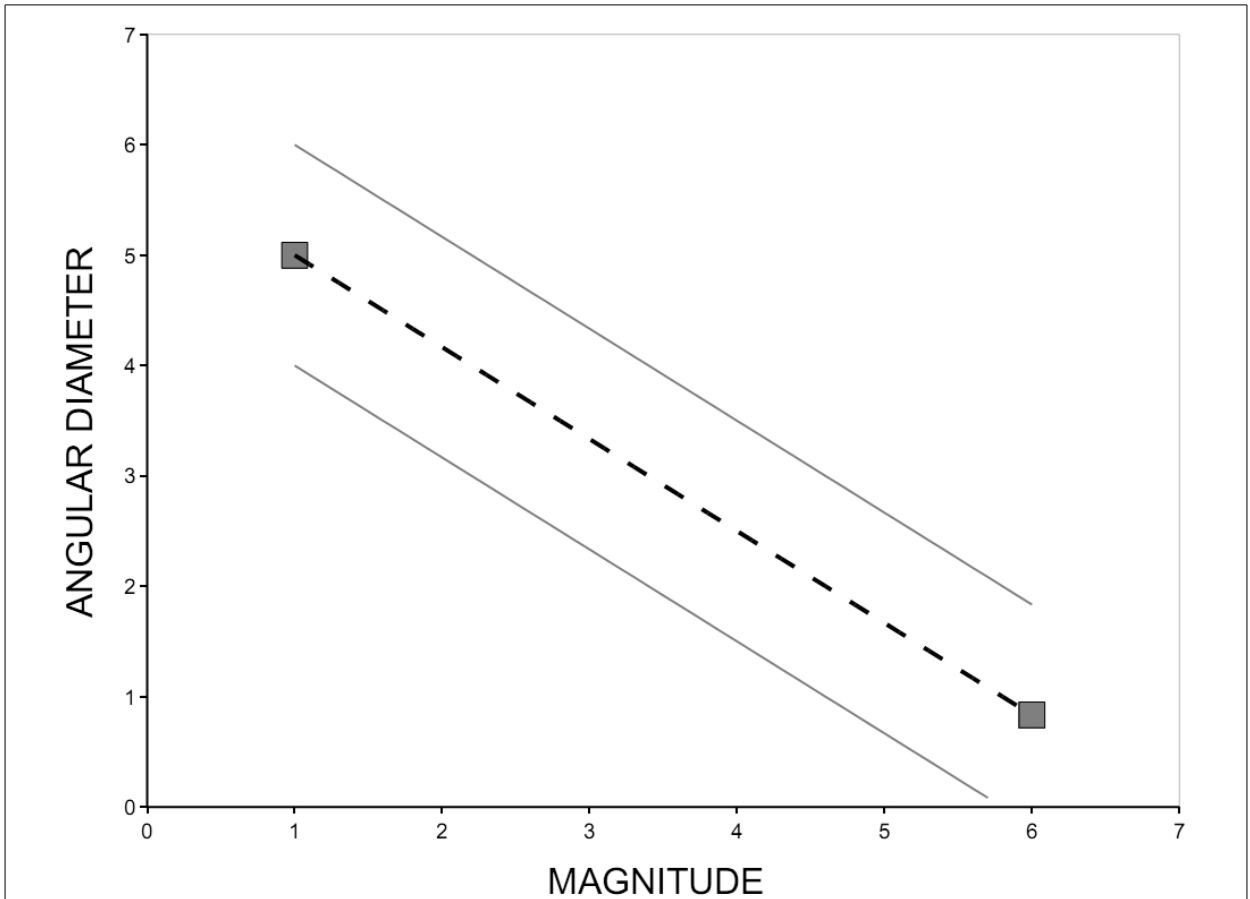

FIGURE 1: Galileo's values of magnitude *M* and angular size $\alpha$ from the *Dialogue* (marked points). Dotted line is linear relationship between *M* and $\alpha$. Thin solid lines represent Galileo's measurement error (see Addendum).

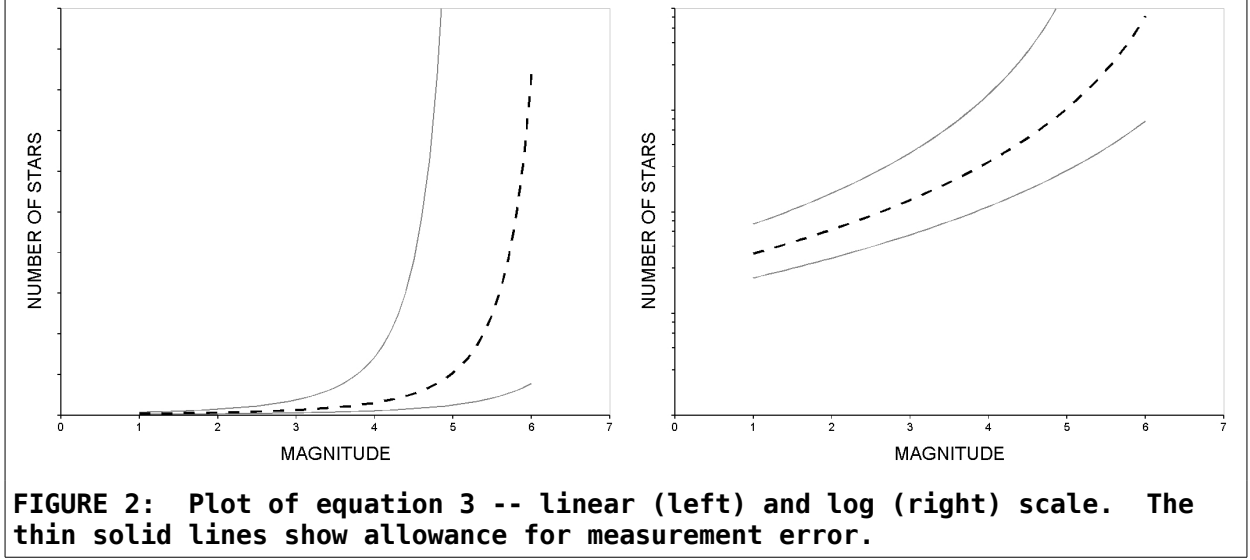

FIGURE 2: Plot of equation 3 -- linear (left) and log (right) scale. The thin solid lines show allowance for measurement error.

would support Galileo's ideas regarding the stars being suns scattered through space.

Data on the numbers of visible stars brighter than a given magnitude are given in Figure 3.[10]  Based on Galileo's assessment that 6$^{th}$ magnitude stars lie at a distance of *2160 AU*, *ρ\* = 1.991 x 10$^{-7}$ star/AU$^3$*.[§‡]  This value used with equation 3 to plot *N\** vs. *M* yields a result consistent with the data (Figure 4).

Galileo did not have access to modern data, but could obtain estimates by counting naked eye stars by magnitude in various sections of the sky and extrapolating to numbers for the sky as a whole.  Some such data was already available -- Ptolemy's *Almagest* contained a catalog of over a thousand stars.

**Conclusion**

Following Galileo's ideas about the stars leads to the conclusion that the aggregate appearance of the naked-eye stars is data in support of those ideas (Figure 5).  To argue against Galileo's ideas -- to say that the stars are *not* suns scattered through space -- requires explaining why it happens to be that *N\** increases with magnitude in a way so consistent with Galileo's ideas.  (Of course this is not direct proof of a

---

§‡ 8404 stars within a radius of *2160 AU* means the stars are contained within a spherical volume of *42.2 x 10$^9$ AU$^3$*, and

*ρ\* = 8404/42.2 x 10$^9$ AU$^3$ = 1.991 x 10$^{-7}$ star/AU$^3$*.

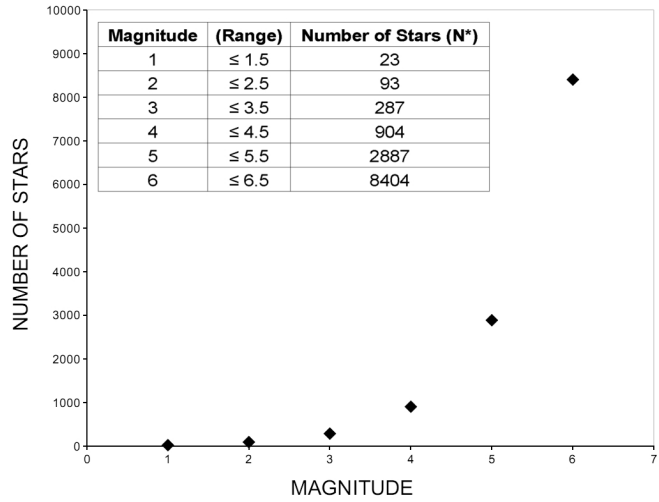

**FIGURE 3:** Table and plot of numbers of stars and magnitudes from *Bright Star Catalog*.

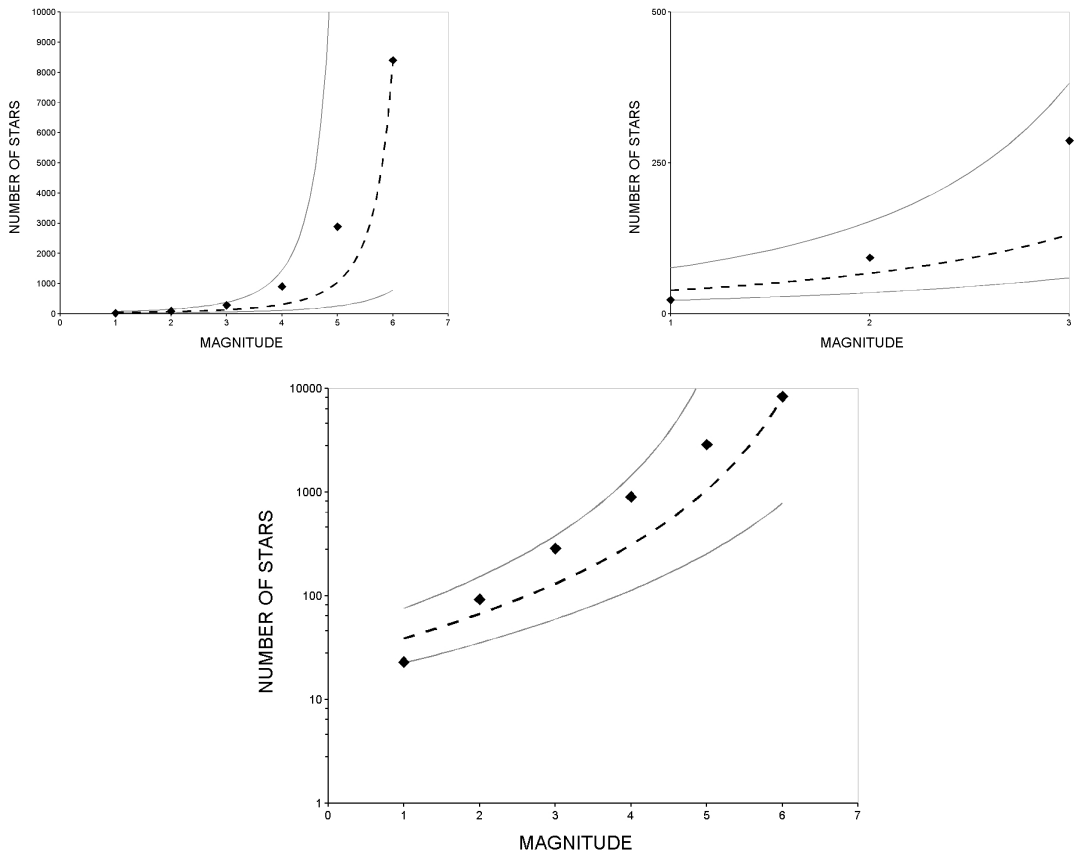

**FIGURE 4:** Data from *Bright Star Catalog* plotted with equation 3 -- linear (top left and right) and log (bottom) scale. The thin solid lines show allowance for measurement error.

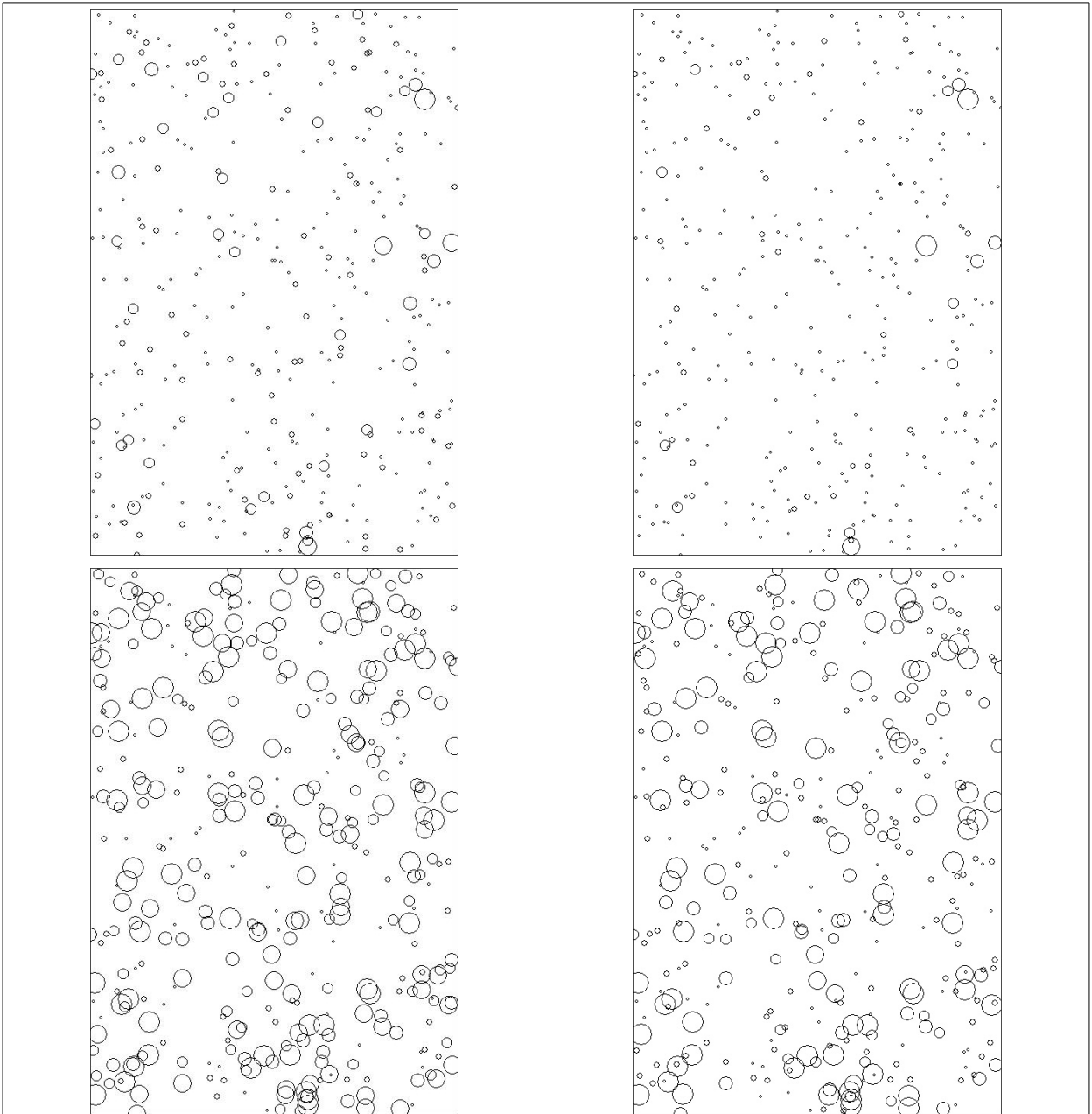

FIGURE 5: Simulated field of stars of magnitudes 1 through 6 (larger circles representing brighter stars). Top left -- numbers of each magnitude in proportions found in *Bright Star Catalog* (i.e. real sky). Top right -- numbers calculated via equation 3. Bottom left, equal numbers of each magnitude. Bottom right -- numbers of each magnitude selected at random. If stars are not suns scattered through space then there is no reason for the real sky to look like the top row. For example, if the stars are simply bodies distributed along a spherical shell centered on Earth as in geocentric theories then there is no reason why their numbers by brightness might not be equal (so that $N*$ increases linearly with $M$) or even random.

moving Earth -- the stubborn might argue that the Earth simply is at rest at the center of a spherical universe measuring thousands of AU in diameter and containing thousands of suns, all but one of which whirl around the Earth daily and one of which circles Earth slightly differently with an entourage of planets.)  Whether this line of thought ever occurred to Galileo -- and if so, what impact it had on his views -- could be a subject for further research by those with good access to original sources.

**Addendum**

*The reader skeptical that Galileo could realistically measure the stellar sizes he claims should keep in mind that Galileo stated such measurements repeatedly, including the mentioned 1617, 1624, and 1632 statements, plus additional comments in 1624 that "no fixed star subtends even 5 seconds, many not even 4, and innumerable others not even 2"[11] and a measurement of Sirius as being 5" in diameter[12].  Galileo says he could distinguish 5" from 4" from 2", and there is every reason to believe he could -- Galileo recorded a change in the apparent size of Jupiter from 41.5" to 39.25" and his measurements are consistent with modern calculations[13]; Galileo made highly accurate sketches and measurements, even of objects as faint as Neptune, to accuracies of 2"[14,15]; the star sizes Galileo measured and the linear relationship between $\alpha$ and M implied in the* Dialogue *are consistent with modern calculations of what would be seen through Galileo's telescopes.[16]*

1   D. B. Wilson, "Galileo's Religion Versus the Church's Science?", *Physics in Perspective*, vol. 1 (1999), pp. 65 - 84.

2   Robert Hooke, *An Attempt to prove the Motion of the Earth from Observations* (1674).  Also http://www.roberthooke.com.

3   Owen Gingerich, "Truth in Science: Proof, Persuasion, and the Galileo Affair", *Perspectives on Science and Christian Faith*, vol. 55 (June 2003), pp. 80-87.

4   Galileo Galilei, *Dialogue Concerning the Two Chief World Systems -- Ptolemaic & Copernican*, translated by Stillman Drake (Los Angeles:  University of California Press, 1967), p. 327.

5   *Dialogue...*, p. 382.

6   L. Ondra, "A New View of Mizar", *Sky & Telescope* (July 2004), pp. 72-75; http://www.leosondra.cz/en/mizar/ for an extended version with detailed references.

7   Galileo Galilei, "Reply to Ingoli" in Finocchiaro M. A. 1989, *The Galileo Affair -- A Documentary History* (Los Angeles:  University of California Press, 1989), p. 167.

8   *Dialogue...*, p. 359.

9   *Dialogue...*, p. 359.

10  D. Hoffleit, *Bright Star Catalog*, 5th Revised Ed. (1991); http://cdsarc.u-strasbg.fr/viz-bin/Cat?V/50

11  "Reply to Ingoli", p. 174.

12  Galileo Galilei, *Le Opere di Galileo -- Edizione Nazionale Sotto gli Auspicci di Sua Maesta re d'Italia*, ed. A. Favaro; 20 vols (Florence, 1890-1909), Vol. III, Pt 2, p. 878.

13  S. Drake, C. T. Kowal, "Galileo's Sighting of Neptune", *Scientific American*, vol. 243 (1980), p. 76.

14  M. Standish, A. Nobili, "Galileo's Observations of Neptune", *Baltic Astronomy*, vol. 6 (1997), pp. 97-104.

15  C. Graney, "On The Accuracy of Galileo's Observations", *Baltic Astronomy*, vol. 16 (2007), pp. 443-449.

16  Graney (2007), p 447.